\documentclass[showpacs,showkeys,preprintnumbers]{revtex4}%
\usepackage{amsfonts}
\usepackage{epsfig}
\usepackage{amsmath}
\usepackage{graphicx}
\usepackage{dcolumn}
\usepackage{bm}
\usepackage{amssymb,mathptmx}
\usepackage{amssymb}%
\begin{document}
\title{Measurement of the neutron lifetime and inverse quantum Zeno effect}
\author{Francesco Giacosa$^{\text{(a,b)}}$}
\email{fgiacosa@ujk.edu.pl}
\author{Giuseppe Pagliara$^{\text{(c)}}$}
\email{pagliara@fe.infn.it}
\affiliation{$^{\text{(a)}}$Institute of Physics, Jan Kochanowski University, ul.
Swietokrzyska 15, 25-406 Kielce, Poland,}
\affiliation{$^{\text{(b)}}$Institute for Theoretical Physics, Goethe University,
Max-von-Laue-Str.\ 1, 60438 Frankfurt am Main, Germany.}
\affiliation{$^{\text{(c)}}$Dip. di Fisica e Scienze della Terra dell'Universit`a di
Ferrara and INFN Sez. di Ferrara, Via Saragat 1, I-44100 Ferrara, Italy}

\begin{abstract}
  Quantum mechanics predicts that the decay rate of unstable systems could be
  effectively modified by the process of the measurement
of the survival probability. Depending on the intrinsic properties of the unstable system
and the experimental setup for the observation, one could obtain the so
called (direct) quantum Zeno and inverse quantum Zeno effects
corresponding to a slowing down or a speeding up of the decay,
respectively. We argue that the inverse quantum Zeno effect is in principle
detectable at a percent level for the neutron decay in experiments
with trapped ultracold neutrons. Conversely, this effect is basically
undetectable in experiments in which the neutron lifetime is measured by
measuring the decays of beams of neutrons. As a test of our claim, we
propose a simple qualitative correlation between the number of neutrons in the
trap and the neutron lifetime: the larger the number, the faster the
decay. Finally we discuss also the presently available measurements of
the neutron lifetime and address the issue of the possible discrepancy
that has been reported among the results obtained by the different experimental techniques.
\end{abstract}

\maketitle

 \section{Introduction}

In the framework of Quantum Mechanics (QM), it is well established
that the decay of unstable states is from a fundamental point of view
non-exponential, see e.g. Ref. \cite{Fonda:1978dk} for a theoretical
description, Ref.  \cite{Reizen1} for the experimental proof of short
times deviations and Ref.  \cite{Rothe} for the experimental proof of
late times deviations from the exponential decay law.

In particular, the survival probability $p(t)$ at early times after the
preparation of the unstable system scales as $p(t)=1-t\gamma(t)$ where the
decay rate $\gamma(t)$ is itself time dependent and vanishes in the limit
$t\rightarrow0$ (how fast it depends on the interaction leading to the decay
and on the kinematics of the decay, see for instance
\cite{Giacosa:2010br,Giacosa:2012hd,Giacosa:2011xa} for decays within
super-renormalizable and renormalizable field theories, respectively).

Short time deviations from the exponential decay law lead inevitably
to a possibility of modifying the effective decay of the unstable
state by means of observations. In particular the most spectacular
effects are the so called (direct) quantum Zeno effect (QZE)
\cite{Misra:1976by,Koshino:2004rw,Degasperis:1974yr} and inverse
quantum Zeno effect (IZE) \cite{KK1,KK2,FP,facchiprl}, which
correspond to a slowing down or a speeding up of the decay rate of the
observed system, see Ref.  \cite{2001PhRvL..87d0402F} for the
experimental proof of both of them. In those experiments, cold atoms
are trapped in the sites of optical lattices and they could "decay"
via a tunneling process which allow them to escape from the confining
potential.  Moreover, the QZE has been verified also for the case of
Rabi oscillations between atomic levels \cite{Itano:1990zz} by a
dedicated procedure to measure the occupation number of the
levels. In general, it is very difficult to observe these subtle QM
effects since the deviations from the exponential decay law occur only
for a very short time scale after the preparation of the system in its
initial (unstable) state.

However, as discussed in Refs. \cite{KK1,KK2,FP}, the IZE is expected
to be more ubiquitous than the QZE. Namely, the experimental
conditions to observe this effect should be less restrictive than the
ones required to observe the QZE.  In particular, in Ref. \cite{KK2}
it has been suggested that in principle the IZE could be realized even
for nuclear $\beta$ decays.  By following this suggestion, we will
study in this paper whether it is possible to prove the existence of
the IZE in association with one of the most important decay process of
nuclear and particle physics, i.e. the decay of the neutron.

Indeed, the process $n\rightarrow p+e^{-}+\bar{\nu}_{e}$ has been the subject
of many theoretical and experimental investigations 
since the very beginning of
particle physics (including the hypothesis about the very existence of neutrinos).
A precise measurement of the lifetime $\tau_{\mathrm{n}}$ of the neutron
is of primary importance for fundamental physics
\cite{Wietfeldt:2011suo}: $\tau_{\mathrm{n}}$ determines directly the
primordial helium abundance within the big-bang nucleosynthesis and the vector
coupling $g_{V}$ characterizing the weak decays of nuclei is strictly
connected with the $V_{ud}$ matrix element and thus to the unitarity of the
CKM matrix. Also, possible rare decay channels of the neutron can reveal the
existence of dark matter particles, such as mirror particles which mix with
the standard model ones \cite{Berezhiani:2018eds,Berezhiani:2018udo}.

There have been several measurements of $\tau_{\mathrm{n}}$ which can be
classified into two categories: (i) beam measurements, which involve the
propagation of a beam of free neutrons (here, the emitted protons are counted) 
\cite{PhysRevLett.65.289,Byrne_1996,PhysRevLett.91.152302,PhysRevC.71.055502,PhysRevLett.111.222501}%
 and (ii) trap measurements, in which a sample of trapped ultracold neutrons
(UCNs)) is monitored and counted
\cite{PhysRevLett.63.593,PICHLMAIER2010221,PhysRevC.85.065503,KHARITONOV198998,SEREBROV200572,Serebrov:2017bzo}. 
Especially the last class of experiments is very interesting from the
perspective of fundamental QM since, as we will discuss in this paper,
they offer the opportunity to test the occurrence of the IZE in the neutron decay.   

The main message of this paper is indeed to show that the IZE could
indeed play a role in future trap experiments aiming at measuring the
neutron lifetime.  Basically, the idea is that the correlations which
are necessarily present in a sample of UCNs, allow for a much more
efficient way of measuring the survival probability of the neutron at
short times scales.

There is an additional peculiar fact related to the neutron decay that
deserves to be mentioned. To date \cite{Wietfeldt:2018upi}, the averages of
$\tau_{\mathrm{n}}$ as computed by considering all the beam experiments
$\tau_{\mathrm{n}}^{\text{beam}}=888.1\pm2.0$ s and all the trap experiments
$\tau_{\mathrm{n}}^{\text{trap}}=879.37\pm0.58$ s separately show a
$\sim4\sigma$ discrepancy, with the latter being $\sim8$ s shorter than the
former. More precisely, the mismatch reads $\Delta\tau=\tau_{\mathrm{n}%
}^{beam}-\tau_{\mathrm{n}}^{trap}=8.7\pm2.1$ s or, in terms of the ratio of
decay widths, $\Gamma_{n}^{\text{trap}}/\Gamma_{n}^{\text{beam}}
=1.0098\pm0.0024$.

This so-called "neutron decay anomaly", confirmed also recently in
\cite{Pattie:2017vsj}, has been the subject of many theoretical investigations
aiming at explaining this discrepancy. One of the most exciting proposals,
involving beyond standard model physics, is based on a possible new decay
channel of the neutron into dark matter particles $n^{\prime}$ slightly
lighter than the neutron \cite{Fornal:2018eol}. In this interpretation, while
beam experiments (which detect the protons generated by the neutron decays)
measure just the branching ratio for the proton decay channel, trap
experiments can measure the whole width of the neutron and therefore, in
presence of an additional decay channel, the neutron decay width is
necessarily larger (thus leading to a shorter -and in this framework correct-
lifetime). Of course, such an interpretation would have far reaching
consequences for the physics beyond the standard model. The dark matter
interpretation has been however criticized: the existence of such a fermion
would imply that it can be formed in the dense core of neutron stars leading
to a strong softening of the equation of state. It would be difficult in such
a case to explain the existence of neutron stars as massive as $2M_{\odot}$
\cite{McKeen:2018xwc,Baym:2018ljz,Motta:2018rxp}.

Another, more important, problem with this interpretation arises when
comparing with the recent precise measurements on beta decay neutron asymmetry
which are consistent with the lifetime as measured by trap experiments and not
with the one obtained by beam experiments \cite{Czarnecki:2018okw}, thus
suggesting that some systematics affects the beam experiments and a new decay
channel is not needed \cite{Dubbers:2018kgh,Markisch:2018ndu}, see also Ref.
\cite{Belfatto:2019swo}. In particular, in Ref. \cite{Fornal:2018eol}, by
using the PDG average for the axial-vector to vector coupling ratio
$\lambda=-1.2723+\pm0.0023$, it turns out that both lifetime measurements are
compatible with the SM within the error bar of $\lambda$. However the most
recent experiments \cite{Dubbers:2018kgh,Markisch:2018ndu} have measured a
slightly larger value for $\lambda=-1.27641(45)_{stat}(33)_{sys}$ which are
definitely in agreement with trap experiments and rule out
the beam experiments results.

To summarize, the discrepancy between $\tau_{\mathrm{n}}^{beam}$ and
$\tau_{\mathrm{n}}^{trap}$ is most likely due to complicated
systematics in the beam measurements as discussed also in Refs.
\cite{Wietfeldt:2018upi,Wietfeldt:2011suo}. New measurements with the
beam technique would be clearly very important to definitely rule out
any possible discrepancy.

In this context, a quite speculative but nevertheless interesting idea
is worth being here investigated: one might assume that the IZE
already took place in ongoing trap experiments. In this way, the IZE
could offer an explanation of the "neutron decay anomaly", since the
smaller decay width $\tau_{\mathrm{n}}^{trap}$ would be a consequence
of the increased decay rate for this particular experimental setup.

The article is organized as follows: In Sec. II we briefly review the QZE and
IZE by introducing a suitable mathematical formalism describing the effect of
measurement/environment via response functions \cite{KK1,KK2}. This formalism
is then applied to the particular case of the neutron in Sec. III. In Sec. IV
we discuss under which conditions the IZE is sizable enough to be detected in
future (or even ongoing) experiments. Conclusions and outlooks are outlined
in Sec. V.

\section{Brief review of the QZE and IZE}

In this section we present how the QZE and the IZE can be described
theoretically. To this end, we use the results of
Refs. \cite{KK1,KK2,FP}, in which a theoretical model for the process
of measurement, that we shall use in our approach, has been
described.

Let us consider a certain decay process of an unstable
state denoted as `$n$', whose decay width is expressed by the function
$\Gamma(\omega).$ Later, the state $n$ will be identified with the
neutron, but here it is a generic unstable quantum state or
particle. The energy $\omega$ reads $\omega=m-\sum_{j=1}^{N}m_{j},$
where $m_{j}$ are the masses of the $N$ decay products of the state
$n.$ The quantity $\omega$ (and thus the mass $m$) are variables,
since the mass of an unstable state is not fixed. As a consequence
$\omega\geq0$ i.e. is a positive real number.  The on-shell value is
obtained for
$\omega_{\text{on-shell}}\equiv\omega_{n}=m_{n}-\sum_{j=1}^{N}m_{j}$,
where $m_{n}=m_{\text{on-shell}}$; the `on-shell' decay width reads
$\Gamma _{n}=\Gamma(\omega_{n})=\Gamma_{\text{on-shell}}$.  The
on-shell mass and the on-shell width are useful concepts for particles
or unstable states which are "narrow" i.e. if their mass
distribution is peaked at $m_{\text{on-shell}}$ and
$\Gamma_{\text{on-shell}}/m_{\text{on-shell}}<<1$. The neutron falls
within this category of unstable states since it has only a weak
interaction decay channel.

A quite general result is that, in presence of a series of
measurements and/or interactions of the system with the environment,
the effective measured decay width may change according to the
weighted average, see \cite{KK1}:
\begin{equation}
\Gamma^{\text{measured}}(\tau)=\int_{0}^{\infty}f(\tau,\omega)\Gamma
(\omega)d\omega\text{ ,}\label{gamman1}%
\end{equation}
where the parameter $\tau$ parameterizes the time-scale interval
between two subsequent collapses (or -in the spirit of decoherence-
dephasing) of the wave function. The functional form of
$f(\tau,\omega)$ -peaked at $\omega_{n}$ and typically symmetric
w.r.t. $\omega_{n}$ - depends on the details of the performed
measurements and of the whole environment coupled with the unstable
system, but three general constraints are:
\begin{equation}
\int_{0}^{\infty}f(\tau,\omega)d\omega=1\text{ , }f(\tau\rightarrow
\infty,\omega)=\delta(\omega-\omega_{n})\text{ , }f(\tau\rightarrow
0,\omega)=\text{small const .}\label{cond}%
\end{equation}

The first condition in Eq. (\ref{cond}) guarantees the
normalization. As a consequence, in the Breit-Wigner limit (in which
no deviations from the exponential decay occurs) the function
$\Gamma(\omega)=\Gamma_{n}$ is a simple constant, then
\begin{equation}
\Gamma^{\text{measured}}(\tau)=\int_{0}^{\infty}f(\tau,\omega)\Gamma
(\omega)d\omega=\,\Gamma_{n}\text{ ,}%
\end{equation}
independently on the precise form of $f(\tau,\omega)$. Hence, as
expected, neither the QZE nor the IZE takes place. This case is however
unphysical from the theoretical point of view, since a constant decay
width and the corresponding Breit-Wigner distributions are only
approximations (actually excellent approximations for most of the
phenomenology of nuclear and particle physics).

The second condition in Eq. (\ref{cond}) assures that, if the system is
undisturbed, one obtains the `on-shell' decay width
\begin{equation}
\Gamma^{\text{measured}}(\tau\rightarrow\infty)=\Gamma_{n}\text{ .}%
\end{equation}

Finally, the third condition in Eq. (\ref{cond}) implies that, for $\tau$ very
small, $f(\tau\rightarrow0,\omega)$ is a (small) constant, hence
\begin{equation}
\Gamma^{\text{measured}}(\tau\rightarrow0)=(\text{small constant})\int
_{0}^{\infty}\Gamma(\omega)d\omega\,\rightarrow0.\label{qze}%
\end{equation}

This limit corresponds to the case in which the measurement
  occurs so quickly after the preparation of the system that the decay is strongly hindered, namely the famous QZE. In the energy domain, this means
that the decay is governed by the high energy tail of $\Gamma(\omega)$ which, as we will discuss later, must be small when $\omega$ large enough.

The functional form of $f(\tau,\omega)$ depends on which type of measurement
is performed. For instance, for instantaneous ideal measurements performed at
times $0,\tau,$ $2\tau,...$ it reads \cite{KK1}:
\begin{equation}
f(\tau,\omega)=\frac{\tau}{2\pi}\frac{\sin^{2}\left[\tau  \left(  \omega
-\omega_{n}\right)  /2\right]  }{[\tau\left(  \omega-\omega_{n}\right)  /2]^{2}%
}.\label{f1}%
\end{equation}

If, instead, a continuous measurement is considered (continuous dephasing, see
Refs. \cite{KK1,KK2,FP} for details) one obtains the Lorentzian form
\begin{equation}
f(\tau,\omega)=\frac{1}{\pi\tau}\frac{1}{\left(  \omega-\omega_{n}\right)
^{2}+\tau^{-2}}.\label{f2}%
\end{equation}
More in general, the emergence of a response function does not need to be
caused by measurements. Indeed the time scale $\tau$ can be
considered to be associated with the decoherence time for the unstable state under study. Namely,
it is not important if the dephasing is caused by an experiment aiming at
ascertaining if the unstable system decayed or not or by the
complex interaction of the unstable system with the environment. 
Indeed, in many physical cases,
the value of $\tau$ induced by the coupling with the environment is smaller
than the one due to the observation process thus making environmental 
dephasing more efficient than the measurement itself \cite{omnes}. 

There is a subtle and important problem related to Eq. (\ref{gamman1}). In
general, $\Gamma(\omega\rightarrow\infty)=0$ sufficiently fast to guarantee
convergence of $\Gamma^{\text{measured}}(\tau)$ (independently on the use of
Eq. (\ref{f1}) or (\ref{f2})). Yet, in most cases, the function $\Gamma
_{n}(\omega)$ may start to decrease only for very large $\omega,$ Moreover,
when the unstable state is \textquotedblleft created\textquotedblright, there
is a certain energy indetermination $\Delta E$, in turn meaning that one
should consider $\omega$ between the range $\omega_{n}-\Delta E$ and
$\omega_{n}+\Delta E=\omega_{C}$. When repetitive collapses are considered (of
whatever type), the case in which each measurement occurs instantaneously
(ideal case) would correspond to $\Delta E=\infty,$ thus the whole function
$\Gamma(\omega)$ should be considered for the computation of the decay rate. However, each realistic measurement in
between (it does not matter if continuous or not) takes a finite time and therefore also
$\Delta E$ is finite.

In our approach we then modify Eq. (\ref{gamman1}) as it follows:
\begin{equation}
\Gamma^{\text{measured}}(\tau,\Delta E)=\int_{\omega_{n}-\Delta E}^{\omega
_{n}+\Delta E}f(\tau,\omega)\Gamma_{n}(\omega)d\omega\label{gamman2}%
\end{equation}
Of course, if $\omega_{\text{n}}-\Delta E<0,$ one should replace $0$ as
the lower bound of the integral. Note, this discussion is only qualitative,
since -in general- a modification of the integration range should also be
accompanied with a modification of $f(\tau,\omega),$ but this aspect can be
neglected if $\Delta E$ is sufficiently larger than the width of the unstable
state (the normalization $\int_{\omega_{n}-\Delta E}^{\omega_{n}+\Delta E}f(\tau,\omega)d\omega\simeq1$ is guaranteed at a very good level of accuracy).

As a simple example, let us assume that
\begin{equation}
\Gamma_{n}(\omega)\simeq g^{2}\omega^{\alpha}%
\end{equation}
in the energy range $(\omega_{\text{n}}-\Delta E,\omega_{\text{n}}+\Delta E$), with $g$ being a coupling strength. By considering $\Delta E\ll\omega
_{\text{n}}$ and, being the function $f(\tau,\omega)$ centered at $\omega=\omega_{\text{n}}$, it is
easy to see that
\begin{equation}
\Gamma^{\text{measured}}(\tau,\Delta E)>\Gamma_{n}\text{ for }\alpha>1\text{
and }\alpha<0\text{ ,}%
\end{equation}
thus the IZE is realized in such a case. Vice-versa, one finds
\begin{equation}
\Gamma^{\text{measured}}(\tau,\Delta E)<\Gamma_{n}\text{ for }0<\alpha<1\text{
,}%
\end{equation}

which corresponds to the standard QZE. For the limiting cases $\alpha=0$ and
$\alpha=1$ no changes occur, $\Gamma^{\text{measured}}(\tau,\Delta
E)=\Gamma_{n}.$ This simple but rather general result confirms that the
IZE\ is even more common than the QZE.

\section{The IZE and the neutron: general features}

Let us now discuss the IZE for the neutron decay. 
First, we introduce the $Q$ value $\omega=m-m_{p}-m_{e}$,
where we consider $m,$ and thus $\omega,$ as variables. The on-shell value is
$\omega_{\text{on-shell}}=\omega_{n}=m_{n}-m_{p}-m_{e}=0.782333$ MeV (using the values from the PDG
\cite{pdg} and neglecting the error).

Since the aim of this paper is a qualitative discussion of the neutron IZE, we use a simplified formula for the decay function of the
nucleon (the Sargent rule) that allows to understand analytically how this
effect could occur:%

\begin{equation}
\Gamma(\omega)=g_{n}^{2}\omega^{5}\,\,\,\,\mathrm{for}\,\,\,\,\omega
\lesssim\omega_{\text{n}}+m_{\pi}\,,\label{gamman}%
\end{equation}
where $g_{n}\propto g_{V}V_{ud}.$ The on-shell value $\Gamma_{n}=\Gamma
(\omega_{n})=g_{n}^{2}\omega_{n}^{5}=\hslash/888.1$ sec$^{-1}=7.41146\cdot
10^{-25}$ MeV implies that $g_{n}=1.59028\cdot10^{-12}$MeV$^{-2}$ if we use
the bottle results and $g_{n}=1.57465\cdot10^{-12}$MeV$^{-2}$ by using the
trap average. The use of the correct phase space expression for the decay formula ($\omega^{5}$ being the
dominant term, see e.g. \cite{Berezhiani:2018udo} ) would not change the
argument of this paper and is left for future studies.

The behavior of $\Gamma(\omega)$ in Eq. (\ref{gamman}) is valid up to the
opening of the strong interaction threshold at $m_{n}+m_{\pi}.$ Moreover,
modifications at much higher energy for $\omega\sim M_{W}$ (the mass of the
weak interaction bosons) are also expected. Indeed, $\Gamma(\omega)$ should
scale as $\omega$ for $\omega$ much larger than $M_{W}$. For $\omega$ even
larger than GUT and/or Planck scales, one should eventually enter into the
domain in which $\Gamma(\omega\rightarrow\infty)=0.$

On the practical level, we will be interested in the behavior of
$\Gamma(\omega)$ in a range which is much smaller than the first threshold,
hence the behavior in\ Eq. (\ref{gamman}) is sufficient for our purposes.

\begin{figure}[ptb]
\begin{centering}
		\epsfig{file=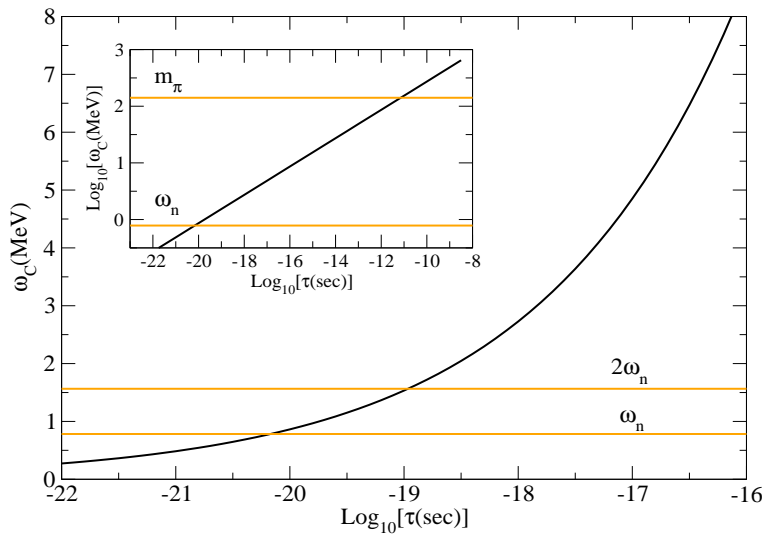,height=10cm,width=7cm,angle=-90}
		\caption{Relation between $\omega_C$ and $\tau$ for a fixed value of the ratio between the decay rate as observed in trap experiments and beam experiments, see text.
The insert displays a wider range of the variables in order to visualize the strong interaction energy scale for $\omega_C$. }
	\end{centering}
\end{figure}

We immediately realize that the use of Eq. (\ref{gamman1}) is problematic in
the case of the neutron: indeed by adopting Eq. (\ref{gamman})
the integral of Eq.
(\ref{gamman1}) is not convergent (if the response functions (\ref{f1}) and (\ref{f2}) are used to model the measurement). Then, the details of
$\Gamma_{n}^{\text{measured}}(\tau)$ would depend also on the high energy
behavior of $\Gamma_{n}(\omega),$ where strong interaction starts to play a
role and -at even higher energies- the effects of the weakly interacting
bosons become important. While this is true in principle, additional
considerations are needed. In practice, it is not realistic to assume that in
the case of the neutron $\omega$ varies up to very large values since the
production processes of the neutron involve much smaller energy uncertainties.

We then turn to the corrected Eq. (\ref{gamman2}). In the case of the neutron,
we assume that $\Delta E \sim$ a few $\omega_{\text{n}},$ since an
off-shellness of few MeV is typical in nuclear processes that produce a
neutron. When applying Eq. (\ref{gamman2}) to the case of the neutron, we
obtain
\begin{equation}
\Gamma^{\text{measured}}(\tau,\omega_{C})=\int_{0}^{\omega_{C}}f(\tau
,\omega)\Gamma_{n}(\omega)d\omega\label{gammaneutronmeas}%
\end{equation}
where $\omega_{C}=\omega_{n}+\Delta E$ is the upper range of the integration.
The lower limit is set to zero since we shall consider $\omega_{C}>\omega_{n}$. Note, for the neutron decay $\alpha=5$
thus the IZE is
clearly favoured according to the general discussion of Sec. II. Moreover, since the integration range in Eq. (\ref{gammaneutronmeas}) is asymmetric
around $\omega_{n}$, the IZE is even more enhanced (as long as Eq. (\ref{gamman})
is used).

For illustrative purposes, we shall use the Lorentzian form of $f(\tau,\omega)$ of Eq. (\ref{f2}). By using Eq. (\ref{gamman2}), the explicit result
reads:
\begin{equation}
\Gamma^{\text{measured}}(\tau,\omega_{C})\simeq\Gamma_{n}\left(  1+\frac
{\hbar}{\tau}\frac{\omega_{C}^{4}}{4\pi\omega_{n}^{5}}\right)
\end{equation}
in the limit of $1/\tau<<\omega_{C}$, which is fulfilled for the decay under
study, see Ref. \cite{KK2}.

As anticipated previously one obtains that:
\begin{equation}
\Gamma^{\text{measured}}(\tau,\omega_{C})>\Gamma_{n}\text{ .}%
\end{equation}
This aspect is \textit{per se} very interesting at a qualitative level: an
increase of the decay width, and hence a decrease of the lifetime, are always
realized. Notice that this result does not depend on the specific choice of
the measurement function $f(\tau,\omega)$: in Ref. \cite{Giacosa:2019akr} it
has been shown that also by using other functional forms, such as the one
describing ideal instantaneous measurements, the results are qualitatively
similar to the ones obtained with the Lorentzian form. This is clear when
considering the limit $1/\tau<<\omega_{C}$, for which both continuous and
instantaneous ideal measurements lead to the very same results.

The important phenomenological question is whether the IZE can be sizable enough to be measured
experimentally. A reasonable estimate to start with shall be $\omega_{C}%
\simeq2\omega_{n}$. Such degree of off-shellness is well below the lowest
threshold for the modifications associated with the strong interaction (and
much below the threshold for weak interaction modifications). Hence,
$\omega_{C}$ is not `intrinsic' to the neutron but is related to the process
of formation and subsequent monitoring of the neutron(s) by the environment
described by the whole physical system, see the discussion in the next subsection.

As a last general point of this section, we discuss why the opposite effect,
the QZE, is excluded for all practical purposes for the case of the neutron.
In fact, in order for Eq. (\ref{qze}) to apply, one needs a very large and
unrealistic value of the parameter $\omega_{C}$ of the order of GUT (or even
larger) scale. In fact, it is necessary for the QZE to occur that, for the
large value of $\omega$ admitted in the integral, the function $\Gamma(\omega)$ is very small. Yet, such a large value of off-shellness can hardly
occur. Moreover, the parameter $\tau$ should be also small enough (much
smaller than $1/M_{W}$ ) to guarantee that such high value of $\omega$
dominate the integral. This discussion shows that in the study of the neutron decay
the QZE can be safely neglected. Similar conclusions have been obtained in Ref. \cite{KK2}. 

\section{IZE in future and ongoing neutron experiments}
\subsection{General considerations}
The main question about the emergence of the IZE in neutron experiments
concerns the value of the parameter $\tau$ that is a measure of how often the neutron
wave function is probed by the environment (not necessarily the `textbook' QM measurement \cite{omnes}).

As a first example, let us take $\omega_{C}$ in the reasonable range $\sim 2$-$10\omega_{n}$ and let us use Eq.~(14) for providing an estimate of
the value of $\tau$ needed to decrease the neutron lifetime 
by a few seconds. It turns out that $\tau$ must be as small
as $10^{-16}$-$10^{-19}$ sec. It is important then to understand if and how
such a small value of $\tau$ can be realized in the case of the neutron lifetime experiments (yet,
similar arguments hold in general also for other weak nuclear decays).

To this end, we now discuss separately both types of experiments for the neutron decay and
we shall see that there are indeed crucial differences between them.

First, let us consider beam experiments. In that case, we should assess how
often does the collapse (or equivalently, the dephasing) of the wave function
take place. Electrons and protons emitted by the neutrons could be quite fast
(typically a velocity of a few tenths of $c$ for the electrons and up to
$\sim10^{-3}c$ for the protons ). Taking into account that the typical
distance that the protons or the electrons have to cover before interacting
with the environment (which includes the detector, but is much larger), thus
causing the collapse, is of about $0.1-1$ m, one obtains that $\tau$ could be
as small as $\sim10^{-9}$ s. For $\omega_{C}$ of the order of $2\omega_{n}$ or
even $10\omega_{n}$, it follows that the beam decay width basically coincides
with the on-shell decay width:%

\begin{equation}
\Gamma^{\text{beam}}\simeq\Gamma^{\text{measured}}(\tau\sim10^{-9}s,\omega
_{C}\sim2\text{-}10\omega_{n})\simeq\Gamma_{n}\text{, }\label{beam}%
\end{equation}
to a very good level of accuracy. In fact, as long as $\omega_{C}$ is
sufficiently small, no deviation from the on-shell value is possible. In fact,
in order to obtain a $\Gamma^{\text{beam}}$ sizably larger than $\Gamma_{n},$
unrealistic values of $\omega_{C}$ are required. For instance, by fixing $\tau=10^{-9}$s and by requiring a $1\%$ IZE: 
$\Gamma^{\text{beam}}\simeq1.01\Gamma_{n}$ one would need $\omega_{C}\sim621\omega_{n}=486$ MeV, which is even sizably larger than the pion mass.
Summarizing, the validity of Eq. (\ref{beam}) seems well upheld. No IZE is
expected in such experiments.

Next, let us turn to the trap experiments. For a single neutron, we also
obtain a similar dephasing time of about $\tau\sim10^{-9}$ s. Yet, here the
situation is different due to fact that trap experiments deal with UCNs
(temperature of the order of 1mK and de Broglie wavelength $\lambda$ greater
than 100nm, see e.g. \cite{Serebrov:2017bzo}), thus neutrons are strongly
correlated (entangled) by the requirement of anti-symmetrization of the wave function.
Moreover the spectrum of neutrons is prepared in the experiment in such a way
that energetic neutrons rapidly escape the trap. Conversely, neutrons in beam
experiments have a rather broad spectrum \cite{Wietfeldt:2011suo} and we
expect therefore to be very weakly correlated to each other. As a consequence,
we can provide this simple qualitative argument:
the collapse of a single neutron (i.e. the detection of one of its decay
products by the environment) implies the collapse of the whole wave function,
and thus of all neutrons. Since one can estimate that there are about $10^{8}$
neutrons in the trap \footnote{The exact number of neutrons in the trap is
obviously not known and indeed the method for measuring the lifetime is based
on the measurement of ratios of number of neutrons for different storage
times. By considering e.g. the experimental setup of \cite{Serebrov:2017bzo},
one can notice that during the initial stage of the preparation of the system,
a flux of the order of $10^{3}$sec$^{-1}$ is measured by the neutron detector
during the filling time which lasts typically a few hundreds seconds. This
leads to a total number of neutrons detected of the order of a few $10^{5}$.
By correcting with the ratio between the surface of the detector and the
surface of the trap we find it reasonable to assume that the trap contains a
factor $10^{2-3}$ more neutrons. This leads to an upper value of $10^{8}$. One
can also extract those estimates by considering the simulated spectra of
neutrons shown in Fig.6 of \cite{Serebrov:2017bzo}.} the effective $\tau$
reads $\tau\sim10^{-9}\cdot10^{-8}=10^{-17}$ s under the assumption of a complete correlation between all the neutron in the trap. If we then consider
\begin{equation}
\Gamma^{\text{trap}}\simeq\Gamma^{\text{measured}}(\tau\sim10^{-17}%
s,\omega_{C}\sim2\text{-}10\omega_{n})\gtrsim\Gamma_{n}\simeq\Gamma
_{n}^{\text{beam}}\text{ ,}%
\end{equation}
$\Gamma^{\text{trap}}$ can be sizable larger than $\Gamma_{n}.$ 
By repeating the previous analysis, in this case with $\tau=10^{-17}$s,
to obtain a $1\%$ IZE, the degree of off-shellness is much smaller:
$\omega_{C} \sim 5$ MeV thus within the reasonable range of values.

The
consequence of this considerations is that the IZE is indeed possible in trap
experiments. Since it is difficult to control the off-shellness parameter
$\omega_{C},$ the most promising way to test this idea is
to decrease the parameter $\tau$ as much as possible.
According to the discussion above, this goal can be achieved
by increasing
the number of neutrons in the traps. In that case, one should measure a
smaller lifetime of the neutron as compared with beam experiments
and one should 
find the following simple qualitative correlation: the larger the number of neutrons in the trap the smaller the lifetime.

\subsection{Discrepancy among neutron lifetime measurements}
As a last topic, we discuss the rather speculative but interesting possibility
that the IZE experiment has been already realized in ongoing trap experiments. In
this way the neutron decay anomaly could be a consequence of the IZE occurring
in trap experiments (and not in the beam ones). For instance, for $\tau
\sim10^{-17}$ and for the quite realistic value $\omega_{C}=6.19\omega_{n}$
one obtains the required value $\Gamma^{\text{trap}}=1.0098\Gamma
^{\text{on-shell}}=1.0098\Gamma^{\text{beam}}.$ In Fig. 1 we display the
relation between $\tau$ and $\omega_{C}.\ $In particular, we show $\omega_{C}$
as a function of $\tau$ for a fixed ratio $\Gamma^{\text{measured}}%
(\tau,\omega_{C}(\tau))=1.0098\Gamma_{n}$. For each value of $\omega_{C},$ one
can read off which value of $\tau$ is needed to obtained a larger decay width,
such as the one found in the trap measurements.

\begin{figure}[ptb]
\begin{centering}
		\epsfig{file=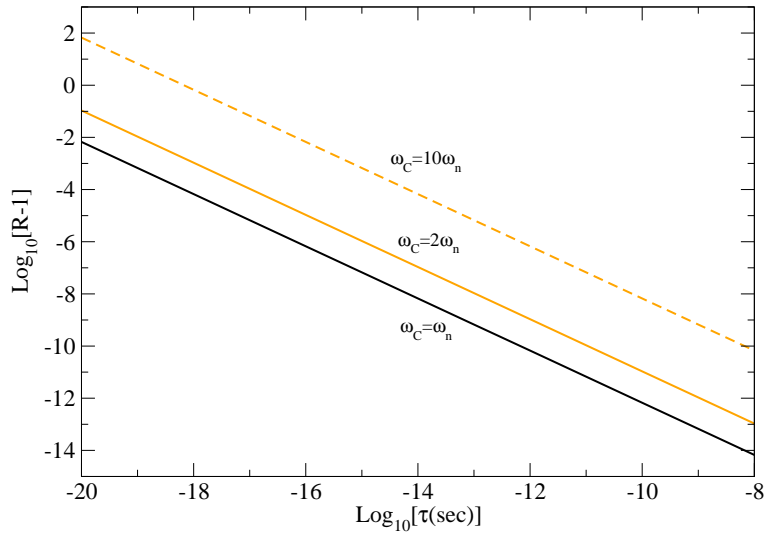,height=10cm,width=7cm,angle=-90}
		\caption{Variation of the ``effective'' decay width as a function of $\tau$ for different values of $\omega_C$. $R=\Gamma_n^{\mathrm{measured}}/\Gamma_n^{\mathrm{on-shell}}$.}
	\end{centering}
\end{figure}

A quite strong implication of this interpretation of the so called "neutron decay anomaly" in
terms of IZE is that the beam experiments provide the `natural' value of the
lifetime of the neutron and in that case a puzzle would arise concerning the
value of $\lambda$ and/or the value of the CKM matrix element $V_{\mathrm{ud}%
}$.

Some kind of `new physics' (but different from an invisible decay as in Ref.
\cite{Fornal:2018eol}) would then be needed, such as violations of CKM
unitarity or nuclear structure effects when extracting the value of
$V_{\mathrm{ud}}$ from superallowed nuclear decays could play a major role,
see discussions in Refs.
\cite{Czarnecki:2019mwq,Seng:2018qru,Gorchtein:2018fxl}. Alternatively, even
if at the present state of knowledge it seems quite improbable, beyond standard model processes
could affect the value of the contribution of the radiative corrections, such
that the lifetime measured in beam experiments is the correct one for an
isolated free neutron. It is then clear that the application of the IZE to
ongoing trap experiments should be regarded with much care, but as long as the
anomaly is not resolved in terms of systematic errors in the beam experiments,
it represents an interesting scenario which is based on
a fundamental quantum phenomenon.

It should be stressed that our mechanism is -at the present stage- only
qualitative, since a very simple measurement model has been used. Yet, it may
point to an interesting possibility that needs further investigation in the
future. As a simple prediction, we note that an additional measurement of
protons in trap experiments should confirm the decreased value of
$\tau^{\text{trap}},$ since this has no influence in our scenario where the
only decay channel of the neutron is the standard beta-decay. Also the
strength of the magnetic field (used in some trap experiments and in beam
experiments) should not change the lifetime of the neutron at variance with
the scenario proposed in \cite{Berezhiani:2018eds} where neutron oscillations
into mirror neutrons are enhanced by the magnetic field. On the other hand, as
already mentioned above, if one could (significantly) decrease or increase the
number of neutrons in the trap, one should correspondingly measure a decrease
or increase the \textquotedblleft effective\textquotedblright\ decay width.
This effect is shown in Fig. 2 where we display the variation of the ratio
$R=\Gamma^{\text{measured}}/\Gamma_{n}$ as a function of $\tau$ for different
values of $\omega_{C}$.

\section{Conclusions}

In this work we have discussed that the inverse quantum Zeno effect,
i.e. the \textit{acceleration} of the decay of unstable systems induced by the measurement, is in principle detectable 
in experiments on the lifetime of the neutron. In particular, trap experiments represent a very interesting setup in which the neutrons are
ultracold and quantum effects may become relevant. Indeed the IZE is induced by a very efficient and fast collapse of the wave function which should occur with time scales of the order of $10^{-17}$s. Such a small time scale is completely out of reach in beam experiments but is instead attainable in trap experiments due to the possibly high degree of entanglement of ultra cold neutrons.
To this end, an increase of the number of neutron in the trap (thus going towards a more degenerate fermionic system)
represents the easiest way to test our proposal, since it should lead to a
smaller measured lifetime of the neutron. 
Quite remarkably, trap experiments on ultracold neutrons decay could probe
the very same quantum effect that has been proven for ultracold atoms trapped in
optical lattices i.e. the inverse quantum Zeno effect
\cite{2001PhRvL..87d0402F}.

In addition, we have also discussed the possibility that the IZE has been
already realized in trap experiments and is the ultimate reason for the
so called "neutron decay anomaly" i.e. the fact that $\tau^{\text{trap }}$ is $1\%$ smaller than $\tau
^{\text{beam  }}$. While in the former case the IZE occurred, in the latter
it cannot be obtained since neutrons are uncorrelated.  
We remark that, presently, the most viable solution to this discrepancy is that
beam experiments are affected by some uncontrolled systematics \cite{Wietfeldt:2011suo}; indeed the value of the neutron lifetime as measured in trap experiments is fully compatible with the most recent measurements of the beta decay neutron asymmetry \cite{Dubbers:2018kgh,Markisch:2018ndu}.

In conclusion, even if the IZE has not yet been observed, future trap experiments on ultra cold neutrons are capable of verifying this very subtle quantum effect and open therefore the possibility of tuning the rate of nuclear decays. Clearly this could have enormous impact also on applied physics and nuclear engineering.

\vskip 0.1cm

\textbf{Acknowledgments}: F.G. thanks S. Mr\'{o}wczy\'{n}ski for useful discussions.

\end{document}